\documentclass[11pt]{article}

\usepackage[preprint]{acl}
\usepackage{enumitem}

\usepackage{times}
\usepackage{latexsym}
\usepackage{ulem}
\usepackage[table]{xcolor}

\usepackage{tabularx}
\usepackage{xltabular}
\usepackage[T1]{fontenc}
\usepackage[utf8]{inputenc}

\usepackage{microtype}

\usepackage{inconsolata}

\usepackage{graphicx}
\usepackage{amsmath}
\usepackage{amssymb}
\usepackage{amsthm}
\usepackage{booktabs}
\usepackage{pifont}
\usepackage{algorithm}
\usepackage{algorithmic}
\usepackage{framed}
\usepackage{tcolorbox}
\tcbuselibrary{breakable}
\usepackage{fancyvrb}
\usepackage{listings}
\usepackage{xcolor}
\usepackage{enumitem}

\title{ColorBrowserAgent: Complex Long-Horizon Browser Agent with Adaptive Knowledge Evolution}

\author{
  \textbf{Jihong Wang}$^{1,*}$,
  \textbf{Jiamu Zhou}$^{1,*}$,
  \textbf{Weiming Zhang}$^{1,2}$,
  \textbf{Teng Wang}$^{1}$,
  \textbf{Weiwen Liu}$^{2}$, \\
  \textbf{Zhuosheng Zhang}$^{2}$, 
  \textbf{Xingyu Lou}$^{1,\dagger}$,
  \textbf{Weinan Zhang}$^{2}$,
  \textbf{Huarong Deng}$^{1}$,
  \textbf{Jun Wang}$^{1}$
  \\
  $^1$OPPO Research Institute \\
  $^2$Shanghai Jiao Tong University \\
  \texttt{wangjihong@oppo.com, zhoujiamu@oppo.com}}

\begin{document}

\maketitle

\begin{abstract}
With the advancement of vision-language models, web automation has made significant progress. 
However, deploying autonomous agents in real-world settings remains challenging, primarily due to site heterogeneity, where generalist models lack domain-specific priors for diverse interfaces, and long-horizon instability, characterized by the accumulation of decision drift over extended interactions.
To address these challenges, we introduce ColorBrowserAgent (\underline{Co}mplex \underline{L}ong-H\underline{or}izon Browser Agent), a knowledge-evolving agent for robust web automation. Our approach addresses these challenges through two synergistic mechanisms: human-in-the-loop knowledge adaptation that transforms sparse human feedback into reusable domain knowledge, and knowledge-aligned progressive summarization that stabilizes long interactions through memory compression.
Extensive experiments on WebArena, WebChoreArena and industrial deployment show that ColorBrowserAgent consistently outperforms strong baselines. It achieves a state-of-the-art success rate of 71.2\% on WebArena and maintains 47.4\% performance under zero-shot transfer setting on WebChoreArena. In commercial deployment, it improves user satisfaction by 19.3\% relatively, verifying its robustness in real-world scenarios.
\end{abstract}

\section{Introduction}

Web browsers serve as the universal central interface for digital workflows across daily user activities and complex enterprise business operations.	%
Recent advances in function-calling capabilities of large vision-language models (VLMs) ~\cite{qwen3,gui_agent_report,gpt4o,pou} have empowered web agents to interpret natural language instructions and directly operate on graphical user interfaces (GUIs).
This capability enables the automation of complex, multi-step browser workflows that traditionally require manual intervention~\cite{gui_survey}.

\begin{figure}[t]
    \centering
    \includegraphics[width=\linewidth]{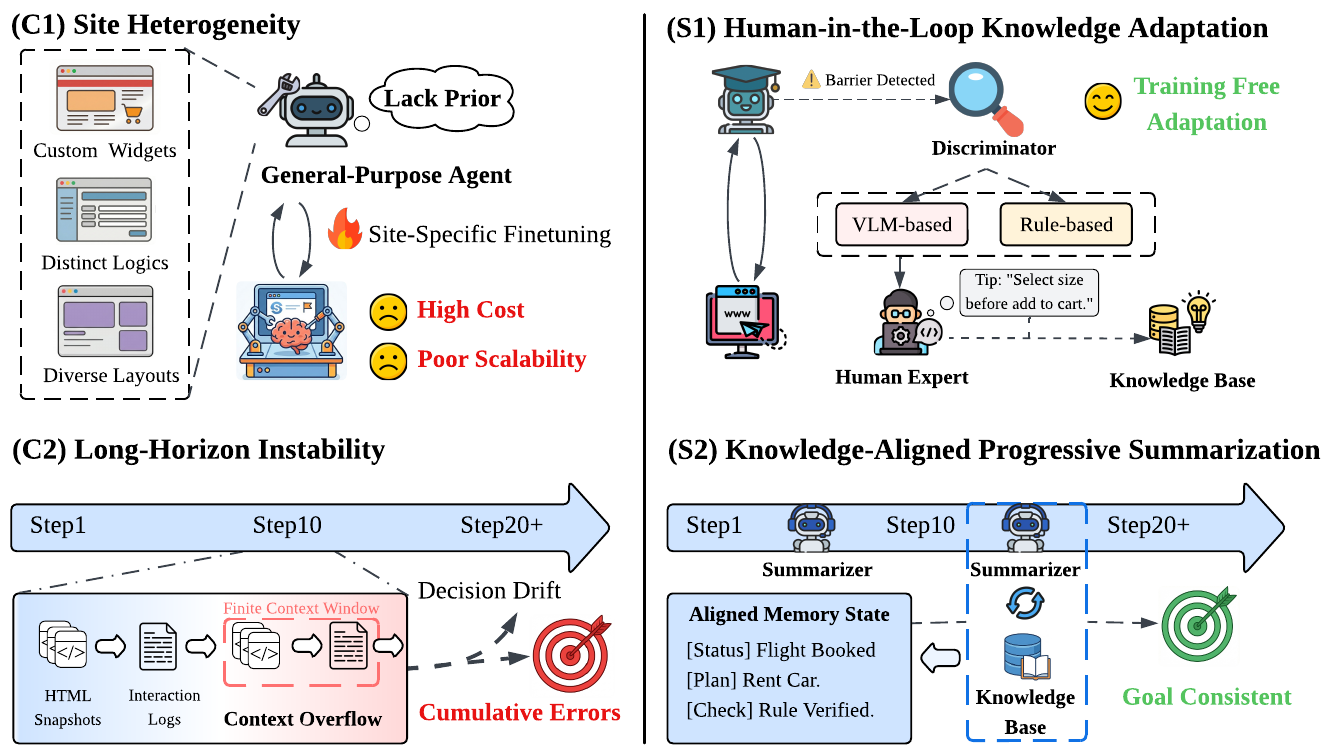}
    \caption{Challenges and Solutions in Complex Web Automation. }
    \label{fig:motivation}
\end{figure}

Despite recent progress, reliable deployment of autonomous web agents in production remains difficult~\cite{illusion_web_agent,reliable_web_agent}. Web interfaces are inherently open and visually driven, introducing substantial variance compared to structured APIs.
In practice, two core challenges impede scalability, as depicted in Figure~\ref{fig:motivation}.
(1) \textbf{Site Heterogeneity.} General-purpose models lack domain-specific priors to handle diverse layout conventions and idiosyncratic business logic across websites. Existing approaches typically rely on post-training using interaction traces ~\cite {webrl,webagent_rl}. This strategy is not only prohibitively expensive but also critically lacks generalization to out-of-distribution websites and robustness to interface evolution.
(2) \textbf{Long-Horizon Instability.} As interaction trajectories extend, agents frequently suffer from decision drift, gradually deviating from the original task objectives~\cite{WebExplorer}. This challenge is exacerbated by site heterogeneity: long-horizon interactions are more prone to lacking domain-specific priors, and minor interface ambiguities (e.g., transient pop-ups) that are negligible in short trajectories can accumulate, leading to severe task deviation.



To address site heterogeneity and long-horizon instability, 
we introduce \textbf{ColorBrowserAgent} 
(\underline{Co}mplex \underline{L}ong-H\underline{or}izon Browser Agent), 
a web agent based on tuning-free adaptation that externalizes domain knowledge and stabilizes long-horizon interaction trajectories. 
As illustrated in Figure~\ref{fig:motivation}, our framework resolves the two challenges through two synergistic mechanisms: 
(1) \textbf{Human-in-the-loop Knowledge Adaptation}, which tackles site heterogeneity by capturing site-specific logic from sparse expert feedback upon failure. These insights are crystallized into an external knowledge base and retrieved contextually, enabling the agent to adapt to diverse interfaces without retraining; and 
(2) \textbf{Knowledge-aligned Progressive Summarization}, which mitigates long-horizon decision drift by maintaining a compressed belief state. This mechanism progressively summarizes interaction history and enforces consistency with retrieved expert priors, ensuring the agent remains focused on the objective throughout long-horizon trajectories.

In summary, our contributions are as follows:

\begin{itemize}
    
\item \textbf{Tuning-Free Adaptation}. 
    We propose a human-in-the-loop knowledge adaptation framework that externalizes domain-specific priors into a persistent memory, decoupling adaptability from model parameters and eliminating the need for retraining.

\item \textbf{Long-Horizon Stability}. 
We introduce a knowledge-aligned progressive summarization mechanism that mitigates decision drift by compressing a belief state across long-horizon interaction trajectories.

\item \textbf{Empirical and Deployment Validation}. 
    ColorBrowserAgent achieves 71.2\% success rate on WebArena and strong zero-shot transfer to WebChoreArena. In a commercial deployment, it further improves user satisfaction by 19.3\%, demonstrating practical effectiveness in dynamic environments.
\end{itemize}

\section{Related Works}

\subsection{LLM-based Web Agents}
Several recent works have introduced realistic browser environments and benchmarks to evaluate LLM-based web agents~\citep{WebArena,survey_web_agent,coloragent,webchorearena}.

Existing approaches fall into two paradigms: \textbf{training-based methods}, which update model parameters with interaction data, and \textbf{test-time search methods}, which improve performance via inference-time exploration. Training-based methods like WebRL ~\cite{webrl} and WebAgent-r1 ~\cite{webagent_rl} learn navigation policies from interaction trajectories via reinforcement learning. However, in real-world industrial settings, obtaining high-quality interaction trajectories is costly and labor-intensive, making frequent retraining infeasible. In contrast, search-based methods (best-first search~\cite{best_first_tree,bpp}, Monte Carlo Tree Search~\cite{WebPilot,hrm}) improve performance by exploring candidate action trajectories at inference time. While they boost task success rates, their extensive inference-time exploration incurs substantial latency and cost, limiting real-time deployment suitability. To address these limitations, ColorBrowserAgent bypasses expensive parameter updates and high-latency search via a collaborative, knowledge-driven mechanism enabling robust, efficient execution.
\subsection{Human-in-the-Loop Strategy}
Human-in-the-Loop (HITL) approaches are widely adopted to improve autonomous agent reliability in complex environments~\cite{HITL_in_AI,HITL_survey,OSCAR}. Existing research focuses on leveraging human data for policy learning. For example, demonstration-based methods like AppAgent~\citep{appagent1, appagent2} collect expert trajectories to initialize agent skills before deployment. Similarly, feedback-driven approaches such as WebCoach~\citep{webcoach} use human corrections to iteratively refine the agent's policy. These methods require significant data accumulation and training overhead to show improvements. In contrast, ColorBrowserAgent treats expert interventions as explicit knowledge injection: instead of model tuning, we encode expert guidance into a retrieval-based knowledge base, enabling immediate failure resolution and cross-task logic reuse without retraining.

\begin{figure*}[t]
    \centering
    \includegraphics[width=\linewidth]{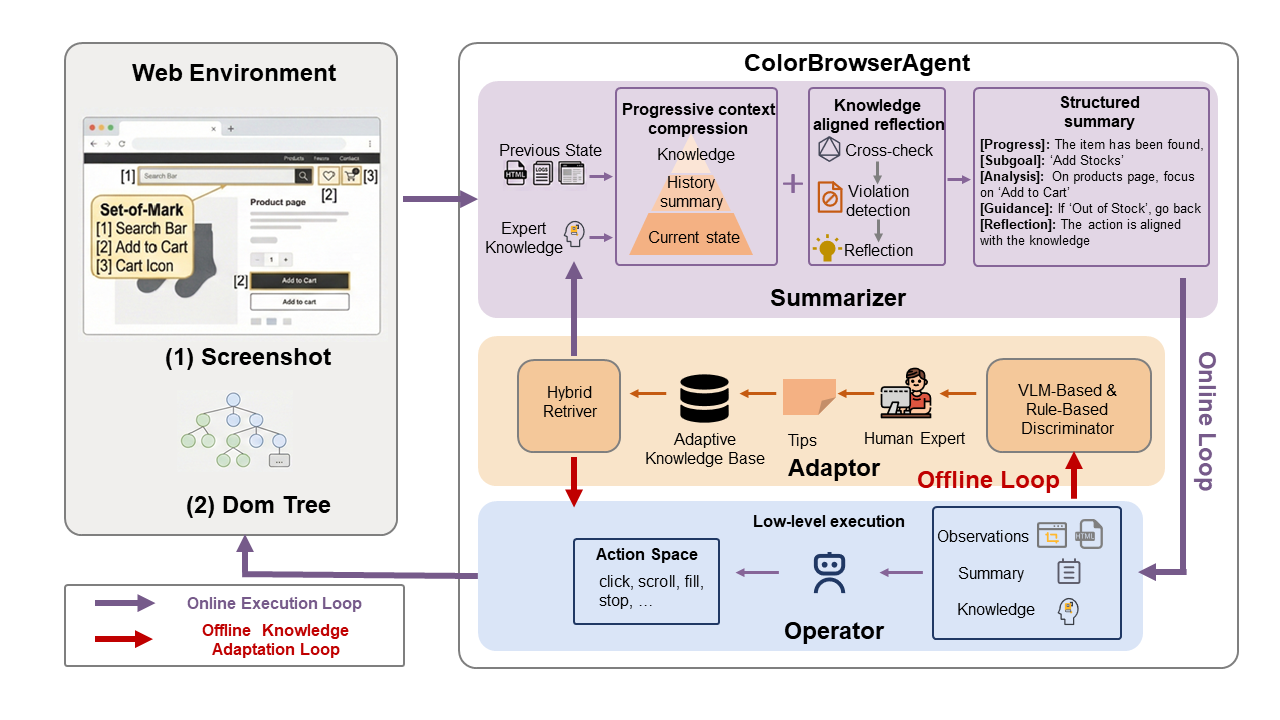}
    \caption{The framework of ColorBrowserAgent. It operates via two synergistic loops: (1) The Offline Knowledge Adaptation Loop, where the Adaptor crystallizes expert feedback into an Adaptive Knowledge Base; and (2) The Online Execution Loop, where the Summarizer maintains a compressed belief state and the Operator executes actions based on aligned intent.}

    \label{fig:framework}
\end{figure*}

\section{Methodology}
\label{sec:method}

\subsection{Preliminary}

We view web automation as a goal-conditioned, sequential decision-making process. 
Given a user instruction $g$, the agent interacts with a browser environment over discrete time steps.
At each time step $t$, the agent receives a multimodal observation $o_t$ consisting of the DOM tree and a rendered screenshot of the current webpage. 
Based on the interaction history $h_t = (o_0, a_0, \dots, o_t)$ and the instruction $g$, the agent selects an action $a_t$ from a predefined action space (e.g., click, type, scroll, stop). A task is considered successful if the terminal environment state satisfies the goal constraints induced by the instruction $g$.

%

\subsection{Overview of ColorBrowserAgent}

Existing approaches~\cite{webagent_rl,webrl} adapt to specific web environments by internalizing site-specific behaviors into model parameters through additional training. However, such parametric adaptation often requires retraining when web interfaces evolve and can struggle to transfer to previously unseen websites. In contrast, we propose \textbf{ColorBrowserAgent}, a tuning-free adaptation framework that externalizes domain-specific expert priors into an explicit knowledge base and retrieves them at runtime to guide action selection.

As shown in Figure~\ref{fig:framework}, ColorBrowserAgent addresses the challenges of site heterogeneity and long-horizon instability through three core components: the Adaptor, the Summarizer, and the Operator. These components function within two synergistic loops:
\begin{itemize}
    \item Offline Knowledge Adaptation Loop: 
To tackle site heterogeneity, the Adaptor constructs the Adaptive Knowledge Base (AKB) through a human-in-the-loop process. 
During an exploratory phase, execution failures are monitored and analyzed by human experts, 
who distill site-specific logic into reusable expert priors and inject them into the AKB.

    \item Online Execution Loop: During inference, the system follows a sequential pipeline: Web Observation $\rightarrow$ AKB Retrieval $\rightarrow$ Summarization $\rightarrow$ Execution. The Summarizer mitigates long-horizon instability by performing knowledge-aligned progressive summarization, compressing interaction history while ensuring consistency with constraints retrieved by the Adaptor. Finally, the Operator generates and executes the concrete browser action.
\end{itemize}
The detailed algorithmic procedure of our framework is provided in Appendix~\ref{sec:appendix_algo}.

\subsection{Adaptor}
The Adaptor fulfills two primary responsibilities: the Offline Knowledge Adaptation Loop for knowledge extraction and injection, and the Online Execution Loop for knowledge retrieval.

\paragraph{Offline Knowledge Adaptation Loop.}
To ensure reliability, the system employs an on-demand intervention strategy via a human-in-the-loop mechanism. It continuously monitors execution via a hybrid detector $f_{\text{trigger}}(o_t, h_t)$, which combines a rule-based detector for deterministic failures and a VLM-based evaluator for semantic inconsistencies between the UI state and the user's intent. 
Intervention is triggered whenever either detector signals a failure. Upon failure detection, a human expert analyzes the failure context and formulates a corrective tip. Crucially, to ensure generalizability, these tips describe the site's operational logic rather than the specific execution flow. This feedback is persistently crystallized into the AKB.

\paragraph{Online Execution Loop.}
During the processing of a new query, the system dynamically retrieves relevant knowledge from the knowledge base via a cascade of mechanisms: (1) URL Pattern Matching for site-specific lookups; (2) Keyword Search for content-aware constraints; and (3) Visual-Semantic Embedding for fuzzy UI matching.

\subsection{Summarizer}

To mitigate decision drift in long-horizon tasks, where the linearly growing history $O(T)$ exhausts the context window and weakens instruction adherence, we introduce a Summarizer agent. 
This supervisory component maintains a structured belief state to preserve temporal coherence and enforce alignment with expert priors. 
The update process iteratively refines this belief state through two core mechanisms:

\paragraph{Progressive Context Compression.} 
To handle extended trajectories, the Summarizer employs a hierarchical retention strategy. It retains fine-grained interaction details only for the current active subgoal, while recursively collapsing completed history into high-level semantic summaries. This mechanism bounds the memory footprint to near-constant size ($O(1)$) regardless of task length, effectively preventing context overflow and the resulting hallucination or loss of goal focus.

\paragraph{Knowledge Aligned Reflection.}
Beyond compression, the Summarizer also serves as a knowledge-aware consistency monitor. 
It compares the agent’s planned actions with the retrieved expert knowledge. 
When a mismatch is detected, 
the Summarizer injects corrective guidance into the observation stream.
This mechanism ensures that the domain priors incorporated by the Adaptor are consistently applied during execution, 
reducing deviations from expert-informed strategies.


\subsection{Operator}

The Operator translates aligned intent into executable browser actions. At each time step, it conditions on the current multimodal observation (DOM tree and screenshot), the retrieved knowledge constraints from the AKB, and the structured summary produced by the Summarizer.
%
Instead of relying on raw interaction history, the Operator utilizes the compressed belief state to focus on the current subgoal while respecting site-level priors. 
It first infers the immediate subgoal implied by the current state, then grounds this intent to specific UI elements by jointly reasoning over DOM attributes and visual cues, similar to recent multimodal GUI agents for web automation.

\section{Experiments}

\begin{table*}[t]
\centering

\caption{Comparison of success rates on the WebArena~\cite{WebArena} benchmark. 
Best results are highlighted in \textbf{bold}, and second-best results are underlined. 
Relative improvements are computed with respect to the strongest baseline. ``-'' means that the result is not reported in the original paper.
}
\label{tab:comparison}

\resizebox{\textwidth}{!}{
\begin{tabular}{lccccccc}
\toprule
\textbf{Method} & \textbf{Overall} & \textbf{Reddit} & \textbf{GitLab} & \textbf{Shopping} & \textbf{Admin} & \textbf{Map} & \textbf{Multisite} \\
\midrule
BrowserGym~\cite{drouin2024workarena} & 15.0 & 20.2 & 19.0 & 17.2 & 14.8 & 25.5 & - \\
LM-TS~\cite{best_first_tree} & 19.2 & 11.3 & 13.9 & 27.8 & 16.5 & 26.6 & 16.7 \\
Go-Browse~\cite{gandhi2025gobrowse} & 22.6 & 30.7 & 15.3 & 22.4 & 25.3 & 17.9 & - \\
AWM~\cite{wang2024agent} & 35.5 & 50.9 & 31.8 & 30.8 & 29.1 & 43.3 & - \\
Branch-n-Browse~\cite{he2025branchnbrowse} & 35.8 & 50.9 & 36.7 & 34.6 & 26.4 & 46.8 & 18.8 \\
WebPilot~\cite{zhang2025bwebpilot} & 37.2 & 65.1 & 39.4 & 36.9 & 24.7 & 33.9 & - \\
AgentOccam~\cite{yang2025agentoccam} & 45.7 & 67.0 & 43.3 & 46.2 & 38.9 & 52.3 & 16.7 \\
AgentSymbiotic~\cite{zhang2025asymbiotic} & 52.1 & 66.0 & 51.0 & 48.0 & 49.0 & \underline{60.0} & 29.0 \\
ScribeAgent~\cite{shen2024scribeagent} & 53.0 & 73.7 & 59.7 & 45.8 & 37.9 & 56.3 & - \\
WebOperator~\cite{dihan2025weboperator} & 54.6 & \underline{76.4} & 52.8 & 49.2 & 55.0 & 55.2 & 31.3 \\
CUGA~\cite{shlomov2025benchmarks} & \underline{61.7} & 75.5 & \underline{61.7} & \underline{58.3} & \underline{62.6} & \textbf{64.2} &  \underline{35.4} \\
\midrule
\textbf{ColorBrowserAgent} & \textbf{71.2} & \textbf{87.4} & \textbf{65.7} & \textbf{72.9} & \textbf{76.4} & 55.9 & \textbf{64.8} \\
\midrule

\rowcolor{green!15}
{\cellcolor{green!15}Relative Improvement}           &15.38\% & 14.38\% & 6.48\% & 25.04\% & 22.04\% &-14.19\% & 83.05\%\\ \hline
\bottomrule

\end{tabular}
}
\end{table*}

\begin{table*}[t]
\centering
\small
\caption{Zero-shot knowledge generalization on WebChoreArena~\cite{webchorearena}. 
Best results are highlighted in \textbf{bold}, and second-best results are underlined. 
Relative improvements are computed with respect to the strongest baseline. 
``--'' indicates that the result is not reported in the original paper.}

\label{tab:webchorearena_results}
\resizebox{0.9\textwidth}{!}{
\begin{tabular}{lccccc}
\toprule
\textbf{Method} & \textbf{Overall} & \textbf{Shopping} & \textbf{Reddit} & \textbf{Admin} & \textbf{GitLab}  \\
\midrule
SteP~\cite{sodhi2023step} & 3.1 & 2.6 & 4.4 & 0.7 & 4.7  \\
BrowserGym~\cite{chezelles2024browsergym}& 21.2  & 15.4 & 15.4 & 26.5 & 27.6  \\
AWM~\cite{wang2024awm}& 22.4   & 18.0 & 14.3 & 30.3 & 26.8 \\
AgentOccam~\cite{yang2024agentoccam} & 21.5  & 21.3 & 11.0 & 30.8 & 22.8 \\
WEBDART~\cite{yang2025webdart}& \underline{31.1} & \underline{35.0} & \underline{26.4} & \underline{33.8} & \underline{29.1}  \\
\midrule
\textbf{ColorBrowserAgent w/o knowledge} & 34.4  & 38.5 & 27.5 & 52.2 & 33.9 \\
\textbf{ColorBrowserAgent}  & \textbf{47.4} & \textbf{43.6} & \textbf{44.0} & \textbf{58.7} & \textbf{53.5} \\
\midrule
\rowcolor{green!15}
{\cellcolor{green!15}Relative Improvement}& 52.41\%  & 24.57\% & 66.67\% & 73.67\% & 83.85\% \\
\bottomrule
\end{tabular}}
\end{table*}



\subsection{Experimental Setup}

We evaluate ColorBrowserAgent on two complementary benchmarks to assess both in-domain performance and zero-shot knowledge generalization:

\begin{itemize}
\item \textbf{WebArena}~\cite{WebArena}: Our primary benchmark, consisting of 812 long-horizon tasks across multiple real-world domains (e.g., Shopping, GitLab, Reddit, Map).
\item \textbf{WebChoreArena}~\cite{webchorearena}: An extended benchmark featuring more complex and memory-intensive tasks. We evaluate in a zero-shot transfer setting, using only knowledge derived from WebArena.

\end{itemize}

We use GPT-5~\cite{gpt5} as the underlying reasoning model. 
Our system is implemented on top of BrowserGym~\citep{BrowserGym} with Playwright for programmatic browser control. 
Observations consist of multimodal inputs, combining Set-of-Marks~\citep{yang2023setofmark} and accessibility tree representations~\citep{WebArena}. 
We adopt an action space adapted from AgentOccam~\citep{yang2024agentoccam} to reduce execution errors. 
The maximum interaction length is set to 30 steps.

We construct the AKB through a one-off exploration phase, where human operators summarize failure cases into reusable site-level tips rather than query-specific traces. 
This process requires less than 1 person-day and yields a compact set of reusable site-level tips totaling 52 distilled rules across domains (GitLab: 13, Map: 7, Reddit: 5, Shopping: 9, Admin: 18).
After construction, the knowledge base remains frozen throughout evaluation, with no test-time human intervention.

\subsection{Performance on WebArena Benchmark}

As detailed in Table~\ref{tab:comparison}, ColorBrowserAgent achieves a 71.2\% overall task success rate on the WebArena benchmark, establishing a new state-of-the-art and substantially outperforming the strongest prior method, CUGA (61.7\%), by a relative margin of 15.38\%. 

A granular analysis across individual domains reveals that our framework delivers consistent, double-digit relative improvements in the majority of categories. Specifically, we observe exceptional gains in domains characterized by diverse interface structures and complex, long-horizon workflows: Multisite (+83.05\%), Shopping (+25.04\%), and Admin (+22.04\%). Additionally, performance in the Reddit domain reaches an impressive 87.4\% success rate. 
These substantial margins suggest that our synergistic approach—combining externalized domain knowledge via the Adaptor and memory stabilization via the Summarizer—effectively mitigates the core challenges of site heterogeneity and long-horizon instability. 
While performance in the Map domain is slightly lower (55.9\%) compared to the best baseline (64.2\%), ColorBrowserAgent still remains highly competitive across the board.



\subsection{Knowledge Generalization on WebChoreArena}

To evaluate whether crystallized site-level priors generalize beyond their original task distribution, 
we directly transfer the frozen knowledge base from WebArena to WebChoreArena without any task-specific adaptation. 
Importantly, no query-level information from WebChoreArena is used during knowledge construction, 
ensuring a strict zero-shot transfer setting.

Despite the substantially increased task complexity in WebChoreArena—characterized by longer reasoning chains and heavier memory demands—ColorBrowserAgent achieves a 47.4\% overall success rate, 
substantially outperforming both the strongest prior method (WEBDART, 31.1\%) and our knowledge-free variant (34.4\%). 
Crucially, the reported results for all baselines on WebChoreArena are directly cited from the WebDART paper~\cite{yang2025webdart} under a GPT-5-based setting, making this comparison backbone-fair. The substantial gain over the strongest baseline demonstrates that the performance improvements cannot be attributed solely to the use of a more advanced LLM.
Furthermore, the performance gap between the knowledge-enabled and knowledge-free variants indicates that reusable site-level priors contribute substantially beyond backbone reasoning alone.

The gains are observed across most domains, with particularly large margins in GitLab and Admin.
This pattern suggests that crystallized knowledge effectively stabilizes long-horizon execution 
while remaining invariant to query complexity. 
Together, these results indicate that domain priors learned from simpler tasks can generalize to substantially more complex scenarios, 
decoupling operational logic from task difficulty.

\subsection{Ablation Study}

\begin{figure}[t]
    \centering
    \includegraphics[width=\linewidth]{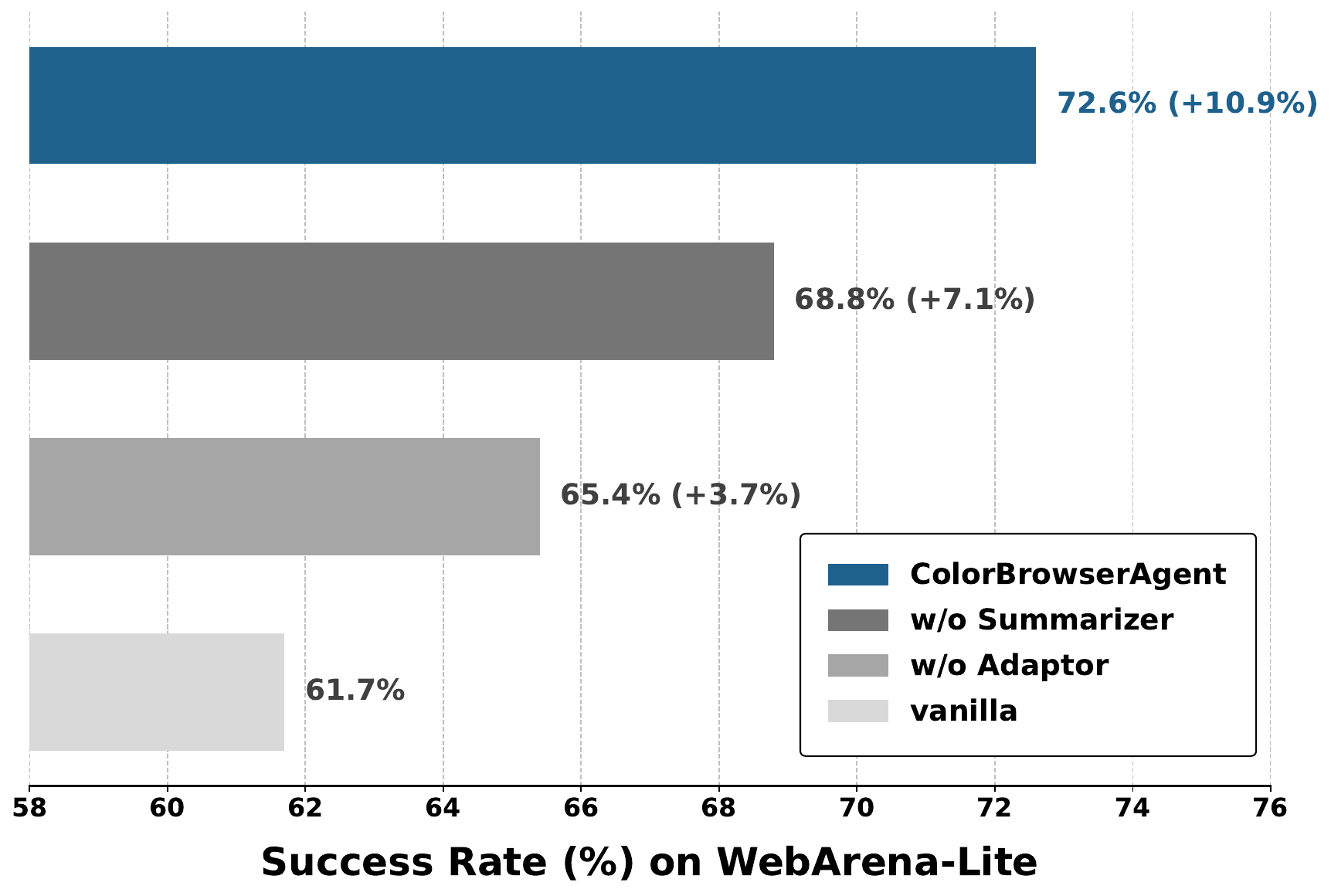}
    \caption{Ablation studies on WebArena-Lite.}
    \label{fig:ablation}
\end{figure}

To assess the contribution of each component, we conduct ablation experiments on WebArena-Lite (165 tasks). Importantly, these experiments use a unified GPT-5 setup, guaranteeing that differences in performance are strictly attributable to the proposed architecture rather than the backbone.
As shown in Figure~\ref{fig:ablation}, the full ColorBrowserAgent achieves a 72.6\% success rate, 
improving over the vanilla baseline by 10.9\%.

Removing Summarizer reduces performance to 68.8\%, indicating that structured memory alignment helps stabilize long trajectories. Removing Adaptor results in a larger drop to 65.4\%, suggesting that externalized site-level priors play a central role in handling heterogeneity.

\subsection{Industrial Practice: Real-Time Service Verification}


We deploy ColorBrowserAgent in a commercial intelligent travel planning system, 
where it operates as the Web Execution Unit within a multi-agent architecture. 
The agent autonomously navigates heterogeneous third-party service portals 
to retrieve real-time information (e.g., flight prices) in response to user requests. 
An online A/B test compares our system against a traditional search-engine-based pipeline.
Note that specific token consumption and cost details are omitted for commercial confidentiality.

\paragraph{Engineering Adaptations.}
To ensure reliability under real-world conditions, we introduce several system-level optimizations. 
First, navigation prioritizes direct URL parameter construction when possible, 
reducing dependence on fragile UI interactions. 
Second, the action space is extended with lightweight primitives such as \texttt{take\_note()} for persistent memory and \texttt{calculate()} for deterministic arithmetic, 
which reduces reasoning errors in multi-step workflows. 
Third, a session watchdog monitors execution and automatically restores state upon transient failures.

\paragraph{Deployment Results.}
In this production environment, the system achieves a 19.3\% relative improvement in the User Satisfaction Rate compared to the baseline. 
Notably, in vertical domains with accumulated site-level priors, the agent exceeds 95\% task success, 
demonstrating that crystallized knowledge remains effective under dynamic interface updates.

\section{Conclusion}

We presented ColorBrowserAgent, a knowledge-evolving web agent designed for reliable industrial deployment. 
By externalizing site-level expertise through Adaptor and stabilizing long interaction trajectories with Summarizer, 
Our framework addresses two practical bottlenecks in production web automation: interface heterogeneity and long-horizon instability.
Extensive evaluations on WebArena and WebChoreArena demonstrate strong performance and zero-shot knowledge transfer without task-specific retraining. 
More importantly, deployment in a commercial real-time service system shows measurable improvements in user satisfaction in dynamic, frequently changing web environments.
These results suggest that decoupling domain knowledge from model parameters through expert-guided persistent memory provides a scalable, low-maintenance paradigm for industrial web agents.

\section*{Limitations}

\paragraph{Cold-Start Adaptation.}
When deployed to entirely new domains without prior knowledge coverage, the system requires increased human intervention to identify and crystallize site-specific logic. 
During this initial phase, performance may fluctuate until sufficient domain priors are accumulated.

\paragraph{Model Dependency.} Our current implementation relies on state-of-the-art proprietary VLMs (e.g., GPT-5) to guarantee high-precision visual reasoning. Distilling these capabilities into smaller, cost-effective open-weights models for edge deployment is a critical next step.


\bibliography{references}

@article{WebArena,
  author={Zhou, Shuyan and Xu, Frank F. and Zhu, Hao and Zhou, Xuhui and Lo, Robert and Sridhar, Abishek and Neubig, Graham},
  title={WebArena: A Realistic Web Environment for Building Autonomous Agents},
  journal={arXiv preprint arXiv:2307.13854},
  year={2023},
  url={https://arxiv.org/abs/2307.13854}
}

@misc{BrowserGym,
      title={The BrowserGym Ecosystem for Web Agent Research}, 
      author={Thibault Le Sellier De Chezelles and Maxime Gasse and Alexandre Drouin and Massimo Caccia and Léo Boisvert and Megh Thakkar and Tom Marty and Rim Assouel and Sahar Omidi Shayegan and Lawrence Keunho Jang and Xing Han Lù and Ori Yoran and Dehan Kong and Frank F. Xu and Siva Reddy and Quentin Cappart and Graham Neubig and Ruslan Salakhutdinov and Nicolas Chapados and Alexandre Lacoste},
      year={2025},
      journal={arXiv preprint arXiv:2412.05467},
      eprint={2412.05467},
      archivePrefix={arXiv},
      primaryClass={cs.LG},
      url={https://arxiv.org/abs/2412.05467}, 
}

@article{yang2023setofmark,
  title={Set-of-Mark Prompting Unleashes Extraordinary Visual Grounding in GPT-4V}, 
  author={Jianwei Yang and Hao Zhang and Feng Li and Xueyan Zou and Chunyuan Li and Jianfeng Gao},
  journal   = {arXiv preprint arXiv:2310.11441},
  year      = {2023}
}

@article{gui_survey,
  title={Large Language Model-Brained GUI Agents: A Survey}, 
  author={Chaoyun Zhang and Shilin He and Jiaxu Qian and Bowen Li and Liqun Li and Si Qin and Yu Kang and Minghua Ma and Guyue Liu and Qingwei Lin and Saravan Rajmohan and Dongmei Zhang and Qi Zhang},
  journal   = {arXiv preprint arXiv:2411.18279},
  year      = {2024}
}

@article{hrm,
  title={Towards hierarchical multi-step reward models for enhanced reasoning in large language models},
  author={Wang, Teng and Jiang, Zhangyi and He, Zhenqi and Tong, Shenyang and Yang, Wenhan and Zheng, Yanan and Li, Zeyu and He, Zifan and Gong, Hailei and Ye, Zewen and others},
  journal={arXiv preprint arXiv:2503.13551},
  year={2025}
}

@inproceedings{bpp,
  title={Bpp-search: Enhancing tree of thought reasoning for mathematical modeling problem solving},
  author={Wang, Teng and Yu, Wing-Yin and He, Zhenqi and Liu, Zehua and HaileiGong, HaileiGong and Wu, Han and Han, Xiongwei and Shi, Wei and She, Ruifeng and Zhu, Fangzhou and others},
  booktitle={Proceedings of the 63rd Annual Meeting of the Association for Computational Linguistics (Volume 1: Long Papers)},
  pages={821--838},
  year={2025}
}

@inproceedings{reliable_web_agent,
  title={Waber: Evaluating reliability and efficiency of web agents with existing benchmarks},
  author={Kara, Su and Faisal, Fazle and Nath, Suman},
  booktitle={ICLR 2025 Workshop on Foundation Models in the Wild},
  year={2025}
}

@article{illusion_web_agent,
  title={An illusion of progress? assessing the current state of web agents},
  author={Xue, Tianci and Qi, Weijian and Shi, Tianneng and Song, Chan Hee and Gou, Boyu and Song, Dawn and Sun, Huan and Su, Yu},
  journal={arXiv preprint arXiv:2504.01382},
  year={2025}
}

@article{WebExplorer,
  title={WebExplorer: Explore and Evolve for Training Long-Horizon Web Agents}, 
  author={Junteng Liu and Yunji Li and Chi Zhang and Jingyang Li and Aili Chen and Ke Ji and Weiyu Cheng and Zijia Wu and Chengyu Du and Qidi Xu and Jiayuan Song and Zhengmao Zhu and Wenhu Chen and Pengyu Zhao and Junxian He},
  journal   = {arXiv preprint arXiv:2509.06501},
  year      = {2025}
}

@article{dihan2025weboperator,
  title={WebOperator: Action-Aware Tree Search for Autonomous Agents in Web Environment}, 
  author={Mahir Labib Dihan and Tanzima Hashem and Mohammed Eunus Ali and Md Rizwan Parvez},
  journal   = {arXiv preprint arXiv:2512.12692},
  year      = {2025}
}

@article{webcoach,
  title={WebCoach: Learning from Human Feedback via Cross-Session Memory Guidance},
  author={Liu, Xiao and others},
  journal={Nature Machine Intelligence (In Press)},
  year={2025}
}

@inproceedings{drouin2024workarena,
  title={Workarena: How capable are web agents at solving common knowledge work tasks?},
  author={Drouin, A. and Gasse, M. and Caccia, M. and others},
  booktitle={Forty-first International Conference on Machine Learning},
  year={2024}
}

@article{best_first_tree,
  title={Tree search for language model agents},
  author={Koh, J. Y. and McAleer, S. and Fried, D. and others},
  journal={arXiv preprint arXiv:2407.01476},
  year={2024}
}

@article{gandhi2025gobrowse,
  title={Go-browse: Training web agents with structured exploration},
  author={Gandhi, A. and Neubig, G. and Dihan, M. L.},
  journal={arXiv preprint arXiv:2506.03533},
  year={2025}
}

@article{wang2024agent,
  title={Agent workflow memory},
  author={Wang, Z. Z. and Mao, J. Y. and Fried, D. and others},
  journal={arXiv preprint arXiv:2409.07429},
  year={2024}
}

@article{he2025branchnbrowse,
  title={Branch-and-browse: Efficient and controllable web exploration with tree-structured reasoning and action memory},
  author={He, S. and Cui, Y. and Ma, X. and others},
  journal={arXiv preprint arXiv:2510.19838},
  year={2025}
}

@inproceedings{zhang2025bwebpilot,
  title={Webpilot: A versatile and autonomous multi-agent system for web task execution with strategic exploration},
  author={Zhang, Y. and Ma, Z. and Ma, Y. and others},
  booktitle={Proceedings of the AAAI Conference on Artificial Intelligence},
  volume={39},
  pages={23378--23386},
  year={2025}
}

@inproceedings{yang2025agentoccam,
  title={Agentoccam: A simple yet strong baseline for llm-based web agents},
  author={Yang, K. and Liu, Y. and Chaudhary, S. and others},
  booktitle={The Thirteenth International Conference on Learning Representations},
  year={2025}
}

@article{zhang2025asymbiotic,
  title={Symbiotic cooperation for web agents: Harnessing complementary strengths of large and small llms},
  author={Zhang, R. and Qiu, M. and Tan, Z. and others},
  journal={CoRR},
  year={2025}
}

@article{shen2024scribeagent,
  title={Scribeagent: Towards specialized web agents using production-scale workflow data},
  author={Shen, J. and Jain, A. and Xiao, Z. and others},
  journal={arXiv preprint arXiv:2411.15004},
  year={2024}
}

@misc{OSCAR,
      title={OSCAR: Optimization-Steered Agentic Planning for Composed Image Retrieval}, 
      author={Teng Wang and Rong Shan and Jianghao Lin and Junjie Wu and Tianyi Xu and Jianping Zhang and Wenteng Chen and Changwang Zhang and Zhaoxiang Wang and Weinan Zhang and Jun Wang},
      year={2026},
      eprint={2602.08603},
      archivePrefix={arXiv},
      primaryClass={cs.AI},
      url={https://arxiv.org/abs/2602.08603}, 
}

@article{gui_agent_report,
  title={Ui-tars-2 technical report: Advancing gui agent with multi-turn reinforcement learning},
  author={Wang, Haoming and Zou, Haoyang and Song, Huatong and Feng, Jiazhan and Fang, Junjie and Lu, Junting and Liu, Longxiang and Luo, Qinyu and Liang, Shihao and Huang, Shijue and others},
  journal={arXiv preprint arXiv:2509.02544},
  year={2025}
}

@article{gpt4o,
  title={Gpt-4 technical report},
  author={Achiam, Josh and Adler, Steven and Agarwal, Sandhini and Ahmad, Lama and Akkaya, Ilge and Aleman, Florencia Leoni and Almeida, Diogo and Altenschmidt, Janko and Altman, Sam and Anadkat, Shyamal and others},
  journal={arXiv preprint arXiv:2303.08774},
  year={2023}
}

@article{qwen3,
  title={Qwen3 technical report},
  author={Yang, An and Li, Anfeng and Yang, Baosong and Zhang, Beichen and Hui, Binyuan and Zheng, Bo and Yu, Bowen and Gao, Chang and Huang, Chengen and Lv, Chenxu and others},
  journal={arXiv preprint arXiv:2505.09388},
  year={2025}
}

@article{webchorearena,
  title={WebChoreArena: Evaluating web browsing agents on realistic tedious web tasks},
  author={Miyai, Atsuyuki and Zhao, Zaiying and Egashira, Kazuki and Sato, Atsuki and Sunada, Tatsumi and Onohara, Shota and Yamanishi, Hiromasa and Toyooka, Mashiro and Nishina, Kunato and Maeda, Ryoma and others},
  journal={arXiv preprint arXiv:2506.01952},
  year={2025}
}

@article{sodhi2023step,
  title={SteP: Stacked LLM policies for web actions},
  author={Sodhi, Paloma and Branavan, SRK and Artzi, Yoav and McDonald, Ryan},
  journal={arXiv preprint arXiv:2310.03720},
  year={2023}
}

@article{chezelles2024browsergym,
  title={The BrowserGym ecosystem for web agent research},
  author={Chezelles, De and Le Sellier, Thibault and Shayegan, Sahar Omidi and Jang, Lawrence Keunho and Lu, Xing Han and Yoran, Ori and Kong, Dehan and Xu, Frank F and Reddy, Siva and Cappart, Quentin and others},
  journal={arXiv preprint arXiv:2412.05467},
  year={2024}
}

@article{wang2024awm,
  title={Agent workflow memory},
  author={Wang, Zora Zhiruo and Mao, Jiayuan and Fried, Daniel and Neubig, Graham},
  journal={arXiv preprint arXiv:2409.07429},
  year={2024}
}

@article{yang2024agentoccam,
  title={AgentOccam: A simple yet strong baseline for LLM-based web agents},
  author={Yang, Ke and Liu, Yao and Chaudhary, Sapana and Fakoor, Rasool and Chaudhari, Pratik and Karypis, George and Rangwala, Huzefa},
  journal={arXiv preprint arXiv:2410.13825},
  year={2024}
}

@article{shlomov2025benchmarks,
  title={From Benchmarks to Business Impact: Deploying IBM Generalist Agent in Enterprise Production},
  author={Shlomov, Segev and Oved, Alon and Marreed, Sami and Levy, Ido and Akrabi, Offer and Yaeli, Avi and Str{\k{a}}k, {\L}ukasz and Koumpan, Elizabeth and Goldshtein, Yinon and Shapira, Eilam and others},
  journal={arXiv preprint arXiv:2510.23856},
  year={2025}
}

@article{yang2025webdart,
  title={WebDART: Dynamic Decomposition and Re-planning for Complex Web Tasks},
  author={Yang, Jingbo and Hou, Bairu and Wei, Wei and Chang, Shiyu and Bao, Yujia},
  journal={arXiv preprint arXiv:2510.06587},
  year={2025}
}

@inproceedings{WebPilot,
  title={WebPilot: A versatile and autonomous multi-agent system for web task execution with strategic exploration},
  author={Zhang, Yao and Ma, Zijian and Ma, Yunpu and Han, Zhen and Wu, Yu and Tresp, Volker},
  booktitle={Proceedings of the AAAI Conference on Artificial Intelligence},
  volume={39},
  pages={23378--23386},
  year={2025}
}

@inproceedings{survey_web_agent,
  title={A survey of webagents: Towards next-generation ai agents for web automation with large foundation models},
  author={Ning, Liangbo and Liang, Ziran and Jiang, Zhuohang and Qu, Haohao and Ding, Yujuan and Fan, Wenqi and Wei, Xiao-yong and Lin, Shanru and Liu, Hui and Yu, Philip S and others},
  booktitle={Proceedings of the 31st ACM SIGKDD Conference on Knowledge Discovery and Data Mining V. 2},
  pages={6140--6150},
  year={2025}
}

@article{coloragent,
  title={ColorAgent: Building A Robust, Personalized, and Interactive OS Agent},
  author={Li, Ning and Lin, Qiqiang and Wu, Zheng and Mo, Xiaoyun and Zhang, Weiming and Zhao, Yin and Qu, Xiangmou and Zhou, Jiamu and Wang, Jun and Zheng, Congmin and others},
  journal={arXiv preprint arXiv:2510.19386},
  year={2025}
}

@article{webrl,
  title={Webrl: Training llm web agents via self-evolving online curriculum reinforcement learning},
  author={Qi, Zehan and Liu, Xiao and Iong, Iat Long and Lai, Hanyu and Sun, Xueqiao and Zhao, Wenyi and Yang, Yu and Yang, Xinyue and Sun, Jiadai and Yao, Shuntian and others},
  journal={arXiv preprint arXiv:2411.02337},
  year={2024}
}

@article{webagent_rl,
  title={Webagent-r1: Training web agents via end-to-end multi-turn reinforcement learning},
  author={Wei, Zhepei and Yao, Wenlin and Liu, Yao and Zhang, Weizhi and Lu, Qin and Qiu, Liang and Yu, Changlong and Xu, Puyang and Zhang, Chao and Yin, Bing and others},
  journal={arXiv preprint arXiv:2505.16421},
  year={2025}
}

@inproceedings{appagent1,
  title={Appagent: Multimodal agents as smartphone users},
  author={Zhang, Chi and Yang, Zhao and Liu, Jiaxuan and Li, Yanda and Han, Yucheng and Chen, Xin and Huang, Zebiao and Fu, Bin and Yu, Gang},
  booktitle={Proceedings of the 2025 CHI Conference on Human Factors in Computing Systems},
  pages={1--20},
  year={2025}
}

@article{appagent2,
  title={Appagent v2: Advanced agent for flexible mobile interactions},
  author={Li, Yanda and Zhang, Chi and Jiang, Wenjia and Yang, Wanqi and Fu, Bin and Cheng, Pei and Chen, Xin and Chen, Ling and Wei, Yunchao},
  journal={arXiv preprint arXiv:2408.11824},
  year={2024}
}

@article{HITL_survey,
  title={A survey of human-in-the-loop for machine learning},
  author={Wu, Xingjiao and Xiao, Luwei and Sun, Yixuan and Zhang, Junhang and Ma, Tianlong and He, Liang},
  journal={Future Generation Computer Systems},
  volume={135},
  pages={364--381},
  year={2022},
  publisher={Elsevier}
}

@article{HITL_in_AI,
  title={Human-in-the-loop artificial intelligence},
  author={Zanzotto, Fabio Massimo},
  journal={Journal of Artificial Intelligence Research},
  volume={64},
  pages={243--252},
  year={2019}
}

@article{pou,
  title={PoU: Proof-of-Use to Counter Tool-Call Hacking in DeepResearch Agents},
  author={Ma, SHengjie and Deng, Chenlong and Mao, Jiaxin and Huang, Jiadeng and Wang, Teng and Wu, Junjie and Zhang, Changwang and others},
  journal={arXiv preprint arXiv:2510.10931},
  year={2025}
}

@misc{gpt5,
  title        = {GPT-5 is here},
  author       = {{OpenAI}},
  howpublished = {\url{https://openai.com/gpt-5/}},
  year         = {2025},
  url          = {https://openai.com/gpt-5/}
}

\newpage
\appendix

\section{Benchmark Details}

\paragraph{WebArena.}
WebArena is a reproducible, realistic benchmark for evaluating web-browsing agents that must execute high-level natural-language objectives through real browser interactions. The environment consists of four self-hosted websites covering distinct domains (online shopping, discussion forum, collaborative development, and an admin/CMS site), plus auxiliary tool sites (e.g., maps). It contains 812 long-horizon tasks specified as user objectives. Agents observe the environment through structured representations such as HTML/DOM or an accessibility-tree, and interact via a compound action space that simulates mouse/keyboard and multi-tab browsing. Concretely, the action space includes element-level actions (\texttt{click/hover/type/press/scroll}), tab management (\texttt{new\_tab/tab\_focus/tab\_close}), navigation (\texttt{goto/go\_back/go\_forward}), and a terminal action \texttt{stop[answer]}.

\paragraph{WebChoreArena.}
WebChoreArena extends WebArena with a focus on more realistic and tedious 'web chores' that require longer procedures and tighter state constraints, while keeping the same websites, interaction protocol, and action space so that agents are directly comparable. It contributes 532 newly curated tasks, including cross-site workflows. Both benchmarks emphasize \emph{functional correctness}: success is determined by whether the final state (and/or produced answer) satisfies the task requirements, rather than matching a specific trajectory.

\paragraph{Queries and evaluation.}
Task queries are natural-language objectives written by human annotators to reflect multi-step, compositional workflows. Evaluation uses two complementary mechanisms:
(i) \emph{answer-based} checks for information-seeking tasks, using exact-match / must-include rules and optionally semantic (LLM-based) fuzzy matching; and
(ii) \emph{programmatic} checks for interaction tasks, where a validator inspects the final web/database state via site APIs, database queries, or DOM/JS selectors and verifies goal constraints. A simple unified view is:
\begin{equation*}
\mathrm{Success}=
\begin{cases}
\mathbb{I}\!\left[\mathrm{Match}(\hat{a},a^*)\right], & \text{answer-based}\\
\mathbb{I}\!\left[\mathrm{Validate}(s_T)\right], & \text{programmatic}
\end{cases}
\end{equation*}
where $\hat{a}$ is the agent answer (if any), $a^*$ is the reference, and $s_T$ is the terminal environment state.

\begin{table}[t]
\centering
\small
\begin{tabularx}{\linewidth}{lcX}
\hline
Benchmark & \#Tasks & Focus \\
\hline
WebArena & 812 & General realistic web tasks \\
WebChoreArena & 532 & More tedious/longer chore-like tasks \\
\hline
\end{tabularx}
\caption{Summary of the two benchmarks used in this work.}
\end{table}

\section{Evaluation Protocol and Human Intervention Policy}
\label{app:protocol}

\subsection{AKB Construction and Refinement Workflow}
\label{app:protocol-webarena}
The construction of the Adaptive Knowledge Base (AKB) adheres to a structured, two-phase iterative process designed to minimize human intervention while maximizing agent adaptability:
\begin{enumerate}
    \item \textbf{Exploratory Deployment Phase}: An initial deployment of the base agent is conducted on WebArena tasks without external knowledge support. During this phase, Adaptor autonomously logs execution anomalies and task failures.
    \item \textbf{Knowledge Distillation Phase}: For identified failures, human experts analyze the interaction traces to diagnose root causes (e.g., non-standard navigation logic, latent UI states). These insights are formalized into generalizable \textit{heuristic entries} following the annotation guidelines (Appendix \ref{app:guideline}) and integrated into the AKB. This process is sparse and efficient, requiring minimal expert effort (approximately one person-day for the entire benchmark).
\end{enumerate}

\subsection{Rigorous WebArena Evaluation Protocol}
\label{app:protocol-webarena-final}
To guarantee evaluation strictness and reproducibility, we enforce the following constraints:
\begin{itemize}[leftmargin=10pt]
    \item \textbf{Static Knowledge Base}: The AKB is strictly \textbf{frozen} prior to the final evaluation. No dynamic knowledge injection or modifications are permitted during inference.
    \item \textbf{Autonomous Execution}: All 812 WebArena tasks are executed in a strictly sequential, fully autonomous manner.
    \item \textbf{Absence of Test-Time Intervention}: The agent must rely exclusively on the pre-compiled knowledge within the static AKB. No human-in-the-loop feedback or real-time correction is accessible during the evaluation phase.
\end{itemize}

\subsection{WebChoreArena Transferability Protocol}
\label{app:protocol-webchorearena}
For the WebChoreArena benchmark, we adopt a strict cross-task transfer learning protocol to assess generalization:
\begin{itemize}[leftmargin=10pt]
    \item \textbf{Zero-Shot Exposure}: The agent is strictly precluded from exposure to WebChoreArena tasks during the knowledge acquisition phase. The AKB comprises solely of heuristics derived from WebArena environments.
    \item \textbf{Direct Cross-Task Transfer}: We evaluate the agent's capacity to generalize site-specific priors learned from WebArena to the more complex query formulations in WebChoreArena, without any task-specific fine-tuning or supplementary knowledge curation.
\end{itemize}

\section{AKB Tip Annotation Guideline}
\label{app:guideline}

\subsection{Tip Granularity and Scope}
\label{app:guideline-granularity}
Tips must capture \textbf{site-level operational logic} rather than query-specific execution traces.
\begin{itemize}[leftmargin=10pt]
    \item \textbf{Bad (Query-Specific)}: "Click the blue button to buy the red socks." (Only valid for one query)
    \item \textbf{Good (Site-Level)}: "To add any item to the cart, you must first select a size option, even if only one size is available." (Valid for all purchase tasks on the site)
\end{itemize}

\subsection{Recommended Tip Template}
\label{app:guideline-template}
Annotators are encouraged to follow this structure:
\begin{itemize}[leftmargin=10pt]
    \item \textbf{Scope}: Define when the tip applies (e.g., "For all product pages...").
    \item \textbf{Action}: Describe the required interaction pattern (e.g., "Hover over the menu...").
    \item \textbf{Constraint}: Mention any critical preconditions (e.g., "...before clicking the link.").
    \item \textbf{Goal Alignment}: Explain why this is necessary (e.g., "...to reveal the submenu.").
\end{itemize}

\section{AKB Statistics and Example Tips}
\label{app:akb}

\subsection{Tip Statistics from WebArena Failures}
\label{app:akb-stats}
The final frozen AKB contains a compact set of tips distributed across the domains:

\begin{table}[h]
\centering
\begin{tabular}{lc}
\toprule
\textbf{Domain} & \textbf{Number of Tips} \\
\midrule
GitLab & 13 \\
Map & 7 \\
Reddit & 5 \\
Shopping & 9 \\
Shopping Admin & 18 \\
\bottomrule
\end{tabular}
\caption{Count of crystallized tips per domain in the Adaptive Knowledge Base.}
\label{tab:akb-stats}
\end{table}

\subsection{Representative Tips by Domain}
\label{app:akb-examples}
Below are representative examples of crystallized knowledge:

\textbf{GitLab}:
\begin{itemize}[leftmargin=10pt]
    \item "When asked to make changes to a repository, commit changes to the main branch UNLESS the task specifically asks you to create a branch."
    \item  "To add new members to the project, you can visit project information > members tab and click blue "invite members" button on top right."
\end{itemize}

\textbf{Reddit}:
\begin{itemize}[leftmargin=10pt]
    \item "if you need find a relevant subreddit or forum, you can find the name after clicking "alphabetical" in the "Forum" tab."
    \item "When creating a post/submission, put all content including URLs in the Body field. Leave the URL field empty."
\end{itemize}

\textbf{Shopping Admin}:
\begin{itemize}[leftmargin=10pt]
    \item "When you add a new product in the CATALOG > Products tab, you can click the  downward arrow beside the "Add Product" button to select options like "Simple Product", "Configurable Product", etc. "
    \item "You can generate various reports by using menus in the REPORTS tab. Select REPORTS > "report type", select options, and click "Show Report" to view report."
\end{itemize}

\textbf{Shopping}:
\begin{itemize}[leftmargin=10pt]
    \item "If you need to find information about your previous purchases, you can go My Account > My Orders, and find order by date, order number, or any other available information"
    \item "An order is considered out of delivery if it is marked as "processing" in the order status."
\end{itemize}

\section{Prompt Templates}
\label{app:prompts}

\subsection{Operator Prompt}
\label{app:prompts-main}
The following system prompt guides the low-level execution agent (Operator).

\begin{tcolorbox}[colback=gray!5!white, colframe=gray!75!black, breakable]
\textbf{Role:}
You are a UI Assistant, your goal is to help the user perform tasks using a web browser. You can communicate with the user via a chat, to which the user gives you instructions and to which you can send back messages. You have access to a web browser that both you and the user can see, and with which only you can interact via specific commands.

\textbf{Instructions:}
Review the instructions from the user, the current state of the page and all other information to find the best possible next action to accomplish your goal. Your answer will be interpreted and executed by a program, make sure to follow the formatting instructions.

\textbf{Output Requirements:}
Your response must strictly follow this output format:

\noindent\texttt{<think>}\\
Explain your reasoning step-by-step here. Describe how you analyze the current page state and your plan to accomplish the user's goal.\\
\texttt{</think>}\\
\texttt{<action>}\\
Only a \textbf{SINGLE} action is allowed in this tag, which will be executed by the program.\\
\texttt{</action>}

\textbf{Formatting Rules:}
\begin{itemize}[leftmargin=10pt]
    \item The \texttt{<think>} tag should contain a clear and detailed explanation of your reasoning process, including how you interpret the task, analyze the page, and decide on your next step.
    \item Inside the \texttt{<action>} tag, you MUST provide exactly one valid action string formatted as ``` ```. This represents the single action to be executed. The action must be one of the valid actions from the action space described below.
\end{itemize}
\end{tcolorbox}

\subsection{ColorSummarizer Prompt}
\label{app:prompts-summarizer}
The Summarizer Agent uses the following prompt to compress history, ensure alignment with retrieved knowledge, and provide guidance.

\begin{tcolorbox}[colback=gray!5!white, colframe=gray!75!black, breakable]
\textbf{Role:}
You are a Progress Summarization and Knowledge Alignment Agent. Your role is to:
1. Analyze the current execution context to track progress.
2. Verify alignment between the agent's actions and the provided expert knowledge (tips).
3. Generate a concise, actionable summary to guide the browser automation agent.

\textbf{Inputs Provided:}
\begin{itemize}[leftmargin=10pt]
    \item \texttt{goal}: The ultimate task objective.
    \item \texttt{relevant\_knowledge}: Key tips or site-specific constraints retrieved from the Adaptive Knowledge Base (AKB).(e.g., "On this site, use the search bar instead of category links.")
    \item \texttt{axtree\_txt}: The accessibility tree of the current webpage.
    \item \texttt{screenshot}: A visual rendering of the current webpage.
    \item \texttt{action\_history}: Sequence of actions already performed.
    \item \texttt{previous\_summary}: The progress summary from the previous step.
\end{itemize}

\textbf{Output Requirements:}
Generate a structured summary with exactly three sections:

\begin{itemize}[leftmargin=10pt]
    \item \textbf{Current Progress \& Knowledge Check}
    \begin{itemize}
        \item State clearly which sub-goals have been completed.
        \item Explicitly evaluate if the recent actions align with \texttt{relevant\_knowledge}.
        \item If a specific tip was violated (e.g., agent clicked a link when the tip suggested searching), flag this deviation immediately.
    \end{itemize}

    \item \textbf{Current State Analysis}
    \begin{itemize}[leftmargin=10pt]
        \item Assess whether the current page is on the critical path toward \texttt{goal}.
        \item Identify  relevant interactive elements (buttons, inputs) necessary for the next step.
    \end{itemize}

    \item \textbf{Next-step Guidance} (Conditional Section)
    \begin{itemize}[leftmargin=10pt]
        \item  Include this section if the agent has deviated from the goal OR violated \texttt{relevant\_knowledge}.
        \item Provide 1–2 concrete recommendations.
        \item If available, quote the specific tip from \texttt{relevant\_knowledge} that dictates the correct next step.
    \end{itemize}
\end{itemize}

\textbf{Formatting Rules:}
\begin{itemize}[leftmargin=10pt]
    \item Always include the  Knowledge Check within the Progress section.
    \item If \texttt{relevant\_knowledge} is empty, proceed with standard progress tracking.
\end{itemize}
\end{tcolorbox}

\section{Practical Insights and Future Directions}

The development and deployment of ColorBrowserAgent have yielded valuable insights into the design of human-centred autonomous systems, shedding light on the pragmatic challenges of integrating agents into human workflows.

\textbf{Engineering for Trustworthy Collaboration.}
While novel architectures often capture the spotlight, our experience underscores that rigorous engineering optimization is the bedrock of user trust. Seemingly mundane enhancements—such as robust error handling for network timeouts, atomic action primitives, and defensive navigation strategies—are essential for transforming an agent from a fragile prototype into a reliable digital partner. In a human-centred context, reliability is not just a metric but a prerequisite for adoption; users will only delegate tasks to agents that demonstrate consistent stability in real-world conditions.

\textbf{Symbiotic Human-AI Collaboration.}
Our results validate that Human-in-the-Loop integration is not merely a fallback mechanism but a powerful paradigm for \textbf{Collaborative Autonomy}. By allowing the agent to request help, we create a symbiotic loop: the AI scales execution for standard tasks, while the human provides high-level intuition for edge cases. A critical avenue for future research is optimizing this interaction cost—developing methods to dynamically identify the "minimal viable intervention" so that the system maximizes the leverage of expert priors while minimizing human cognitive load. The ultimate goal is an agent that learns from its user's guidance, progressively requiring less intervention over time.

\textbf{Foundation Models as Reasoning Engines.}
Finally, our findings reaffirm the critical role of strong foundation models in interpreting human intent. The agent's ability to generalize to novel sites relies heavily on the reasoning capabilities of the underlying LLM. We observe that a capable model (like GPT-5) serves as the bridge between abstract human instructions and concrete interface actions. As these models evolve, we envision a future where agents can infer not just "how" to click, but "why" a user prefers a certain workflow, leading to more personalized and intuitive automation.

\section{Algorithms}
\label{sec:appendix_algo}

Algorithm~\ref{alg:framework} presents the unified process of ColorBrowserAgent, encompassing both the offline construction of the Adaptive Knowledge Base and the online execution phase.

\begin{algorithm}[h]
\caption{ColorBrowserAgent: Knowledge Adaptation and Execution}
\label{alg:framework}
\begin{algorithmic}[1]
\REQUIRE Training Tasks $\mathcal{T}_{\text{train}}$, Test Instruction $g_{\text{test}}$, Environment $\mathcal{E}$, Human Expert $\mathcal{H}$
\ENSURE Adaptive Knowledge Base $\mathcal{K}$, Interaction Trajectory $h$

\STATE \textbf{Phase 1: Offline Knowledge Adaptation Loop}
\STATE Initialize $\mathcal{K} \leftarrow \emptyset$
\FORALL{task $g \in \mathcal{T}_{\text{train}}$}
    \STATE Run base agent on $g$, monitor with $f_{\text{trigger}}$
    \IF{failure detected}
        \STATE Expert $\mathcal{H}$ analyzes failure and extracts site-level logic $\tau$
        \STATE $\mathcal{K} \leftarrow \mathcal{K} \cup \{\tau\}$ \COMMENT{Inject into AKB}
    \ENDIF
\ENDFOR

\STATE \textbf{Phase 2: Online Execution Loop (Inference)}
\STATE Initialize history $h_0 \leftarrow \emptyset$, belief state $m_0 \leftarrow \emptyset$
\STATE Target goal $g \leftarrow g_{\text{test}}$
\FOR{$t = 0$ \TO $T_{\max}$}
    \STATE $o_t \leftarrow \mathcal{E}.\text{observe}()$ \COMMENT{Obtain observation}
    
    \STATE \textbf{// Adaptor: Retrieve Knowledge}
    \STATE $k_t \leftarrow \text{Retrieve}(\mathcal{K}, o_t)$ 

    \STATE \textbf{// Summarizer: Update Belief State}
    \STATE $m_t \leftarrow \text{Summarize}(m_{t-1}, o_t, k_t)$ \COMMENT{Align with knowledge}

    \STATE \textbf{// Operator: Execute Action}
    \STATE $a_t \leftarrow \text{Policy}(o_t, m_t, k_t)$ 
    \STATE $\mathcal{E}.\text{step}(a_t)$
    \STATE $h_{t+1} \leftarrow h_t \cup \{o_t, a_t\}$

    \IF{$a_t$ is \texttt{stop}}
        \RETURN $h_{t+1}$
    \ENDIF
\ENDFOR
\end{algorithmic}
\end{algorithm}

\onecolumn
\section{Case Study}
\label{app:case-study}

To intuitively illustrate the execution of ColorBrowserAgent, we present a failure trajectory in the absence of prior knowledge and demonstrate how injecting priors resolves the issue.

\renewcommand{\tabularxcolumn}[1]{m{#1}}

\begin{table*}[h]
\centering
\small
\caption{The failure trajectory of ColorBrowserAgent without prior knowledge.}
\label{tab:fail_case}
\begin{tabularx}{\textwidth}{m{0.45\textwidth} X m{0.08\textwidth}}
\toprule
\multicolumn{3}{p{0.95\textwidth}}{\textbf{Query:} Add a new color option brown to the size S of Phoebe Zipper Sweatshirt} \\
\midrule
\textbf{Screenshot} & \textbf{Thought} & \textbf{Remark} \\
\midrule
\includegraphics[width=\linewidth]{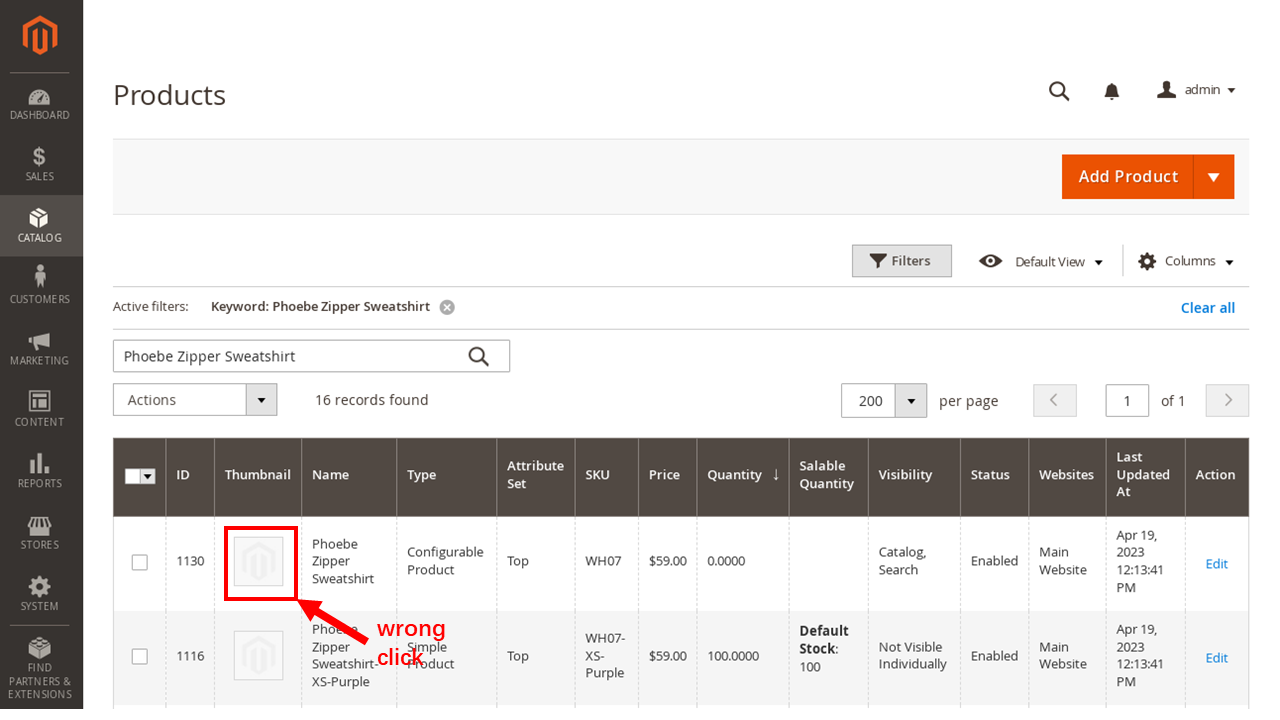} & <think>
The user wants to add a new color option "brown" to size S for the product "Phoebe Zipper Sweatshirt." I’m on the Magento Admin Products grid page with a keyword filter applied for "Phoebe Zipper Sweatshirt," and the grid shows 16 results. To proceed, I need to open the specific product’s edit page. The SOM context lists a grid cell for the Name “Phoebe Zipper Sweatshirt” with bid 1773, which is typically clickable and opens the product edit view. This is the logical next step before accessing configurations to add the new color option for size S. \textbf{I will click the product name cell to open its edit page.}
</think>
 & Navigating to Product. \\
\midrule
\includegraphics[width=\linewidth]{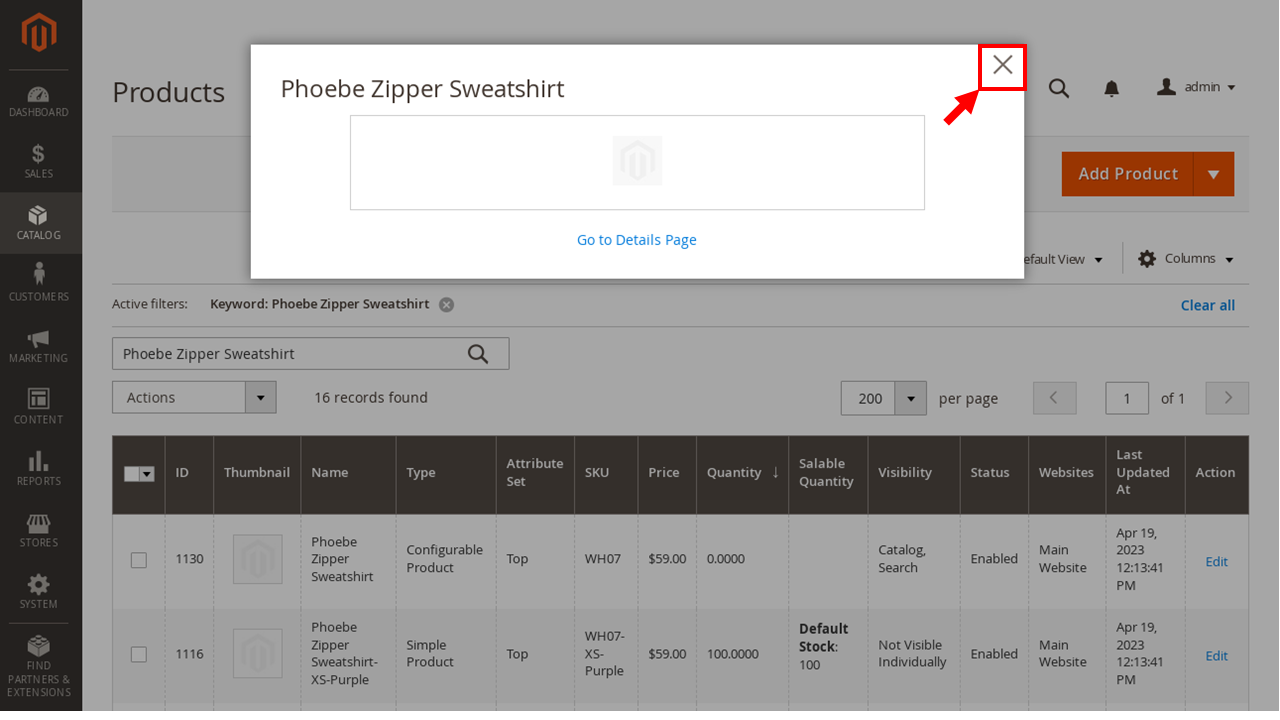} & <think>
The goal is to edit the configurable product “Phoebe Zipper Sweatshirt” to add a new color option brown for size S. I previously clicked the product name in the grid to open its edit page, but the current screenshot still shows the Products grid and a modal dialog overlay near the top (it appears to be a grid-related modal like “Reset Grid Columns”), which likely blocked navigation. \textbf{To proceed, I need to dismiss this modal.} Since I don’t have the specific bid for the modal’s Cancel/Close button, the most reliable way to close it is to press Escape, which commonly dismisses modals in Magento’s admin UI. After the modal is closed, I will be able to click into the product and continue with the configuration changes.
</think>
 & Missed Button. \\
\midrule
\includegraphics[width=\linewidth]{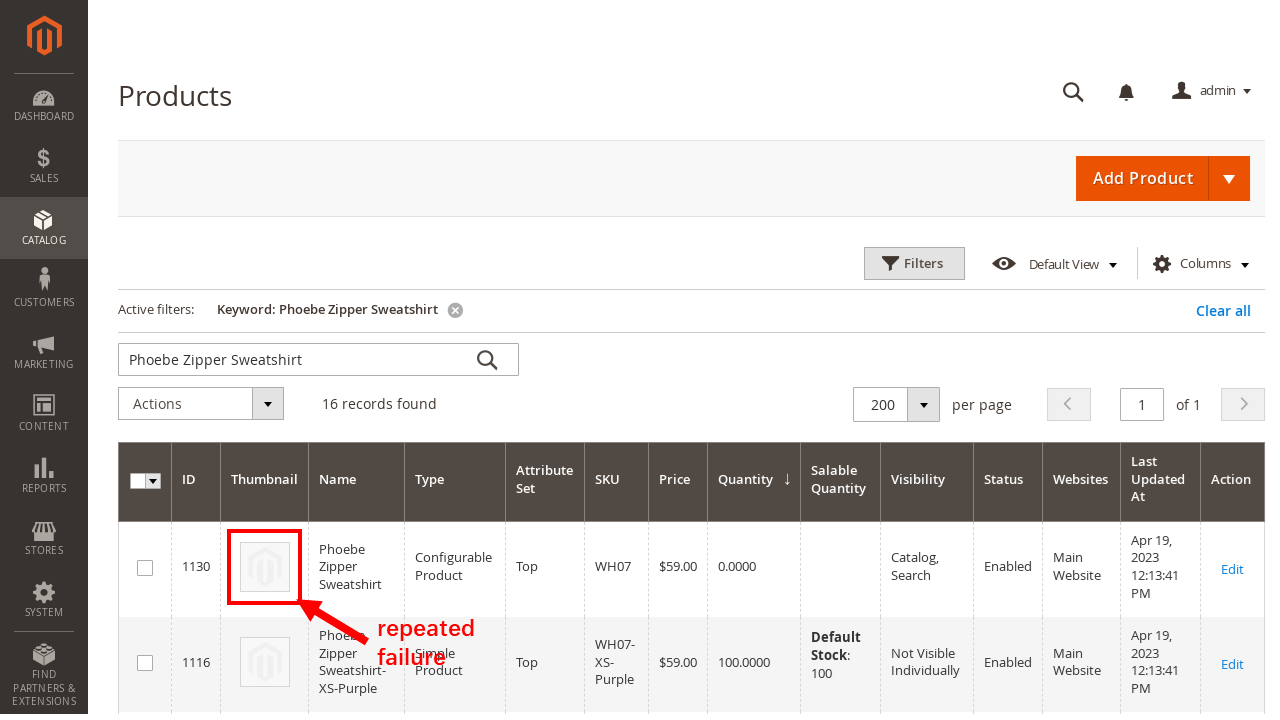} &  <think>
The task is to add a new color option "brown" to size S for the product "Phoebe Zipper Sweatshirt" in Magento. I'm on the Products grid page with the keyword filter applied, and earlier a modal overlay prevented navigation when I tried to open the product. I dismissed the modal by pressing Escape, so the next logical step is to open the product's edit page. The product name cell "Phoebe Zipper Sweatshirt" (bid 1773) is typically clickable to navigate to the edit view for a configurable product. Opening this page is necessary before accessing the Configurations section to add the new color-size variant. \textbf{I will click the product name cell now}.
</think>
 & Trapped in Loop. \\
\bottomrule
\end{tabularx}
\end{table*}

\begin{xltabular}{\textwidth}{m{0.45\textwidth} >{\small}X >{\small}m{0.08\textwidth}}
\caption{The success trajectory of ColorBrowserAgent with injected knowledge.} \label{tab:success_case} \\
\toprule
\multicolumn{3}{p{0.95\textwidth}}{\small \textbf{Query:} Add a new color option brown to the size S of Phoebe Zipper Sweatshirt} \\
\midrule
\multicolumn{3}{p{0.95\textwidth}}{\small \textbf{Knowledge:} If you need to add new attribute values (e.g. size, color, etc) to a product, you can find the product at CATALOG > Products, search for the product, edit product with ``Configurable Product'' type, and \textbf{use ``Edit Configurations'' to add the product with new attribute values}. If the value that you want does not exist, you may need to add new values to the attribute.} \\
\midrule
\textbf{Screenshot} & \textbf{Thought} & \textbf{Remark} \\
\midrule
\endfirsthead
\caption[]{The success trajectory of ColorBrowserAgent with injected knowledge (continued).} \\
\toprule
\multicolumn{3}{p{0.95\textwidth}}{\small \textbf{Query:} Add a new color option brown to the size S of Phoebe Zipper Sweatshirt} \\
\midrule
\textbf{Screenshot} & \textbf{Thought} & \textbf{Remark} \\
\midrule
\endhead
\midrule
\multicolumn{3}{r}{\small \textit{Continued on next page}} \\
\midrule
\endfoot
\bottomrule
\endlastfoot
\small
\includegraphics[width=\linewidth]{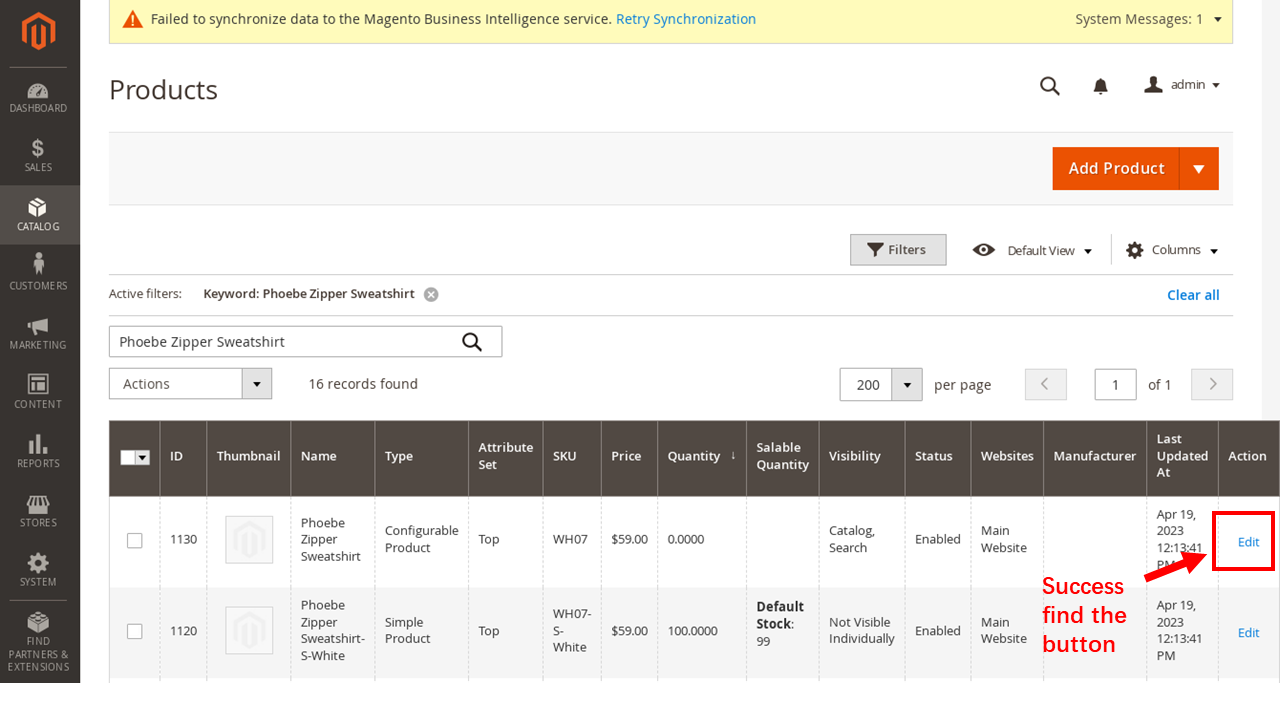} & <think> The task is to add a new color option ``brown'' specifically for size S of the ``Phoebe Zipper Sweatshirt,'' which is a configurable product. On the current Products page, the grid lists the parent configurable product and all its child simples. To modify configurations (add the S-Brown variant), I need to open the parent configurable product's edit page. In the table, I see the row for ``Phoebe Zipper Sweatshirt'' with Type ``Configurable Product'' and \textbf{an Action cell containing the link ``Edit Phoebe Zipper Sweatshirt''}. The correct next step is to click this edit link to access the product editing interface, \textbf{where I can use ``Edit Configurations'' to add the new color option}. If ``brown'' doesn't exist as a Color attribute value, I may need to add it later under STORES > Attributes > Product, but first I must open the product. </think> & Navigating to Product. \\
\midrule
\includegraphics[width=\linewidth]{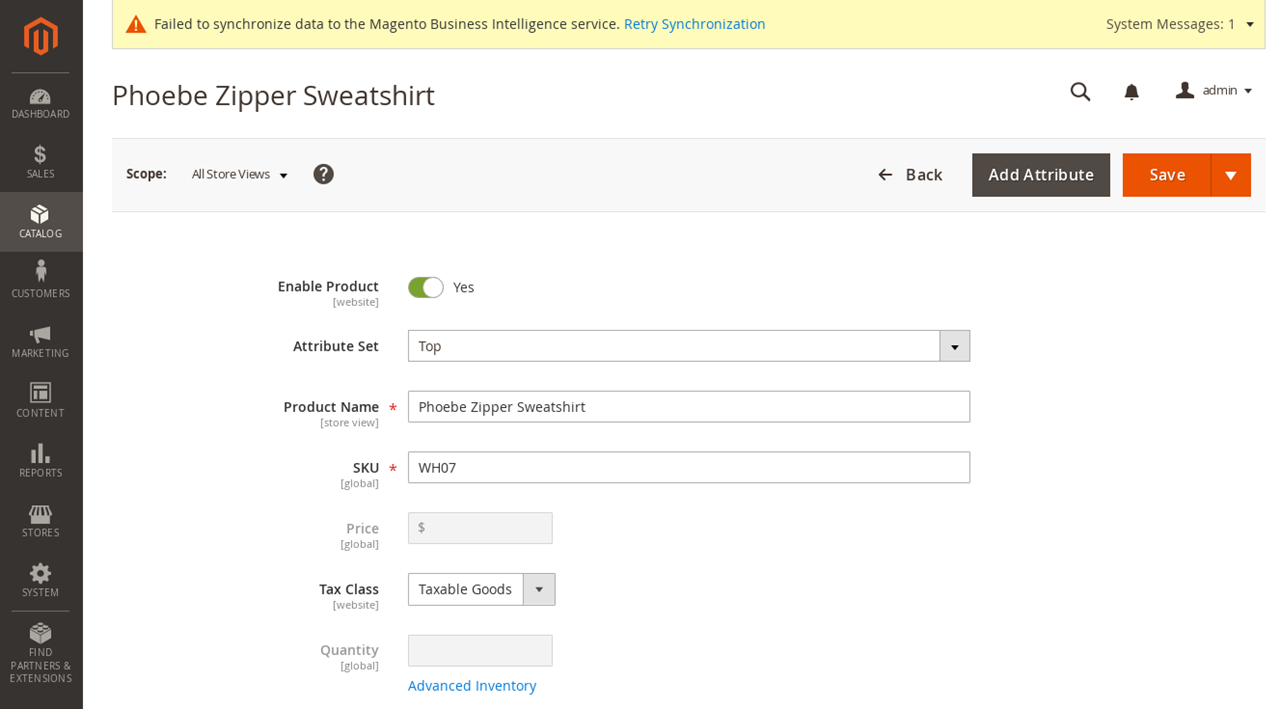} & <think> The goal is to add a new color option Brown to size S for the configurable product ``Phoebe Zipper Sweatshirt.'' I am currently on the product search page and have located the item. The next step is to enter the product edit page. However, the specific configuration button ``Edit Configurations'' is not immediately visible in the current viewport. I need to scroll down the page to locate the ``Configurations'' section where the attribute modification options are typically found. Therefore, the immediate action is to scroll down to reveal the configuration interface. </think> & Locating Config. \\
\midrule
\includegraphics[width=\linewidth]{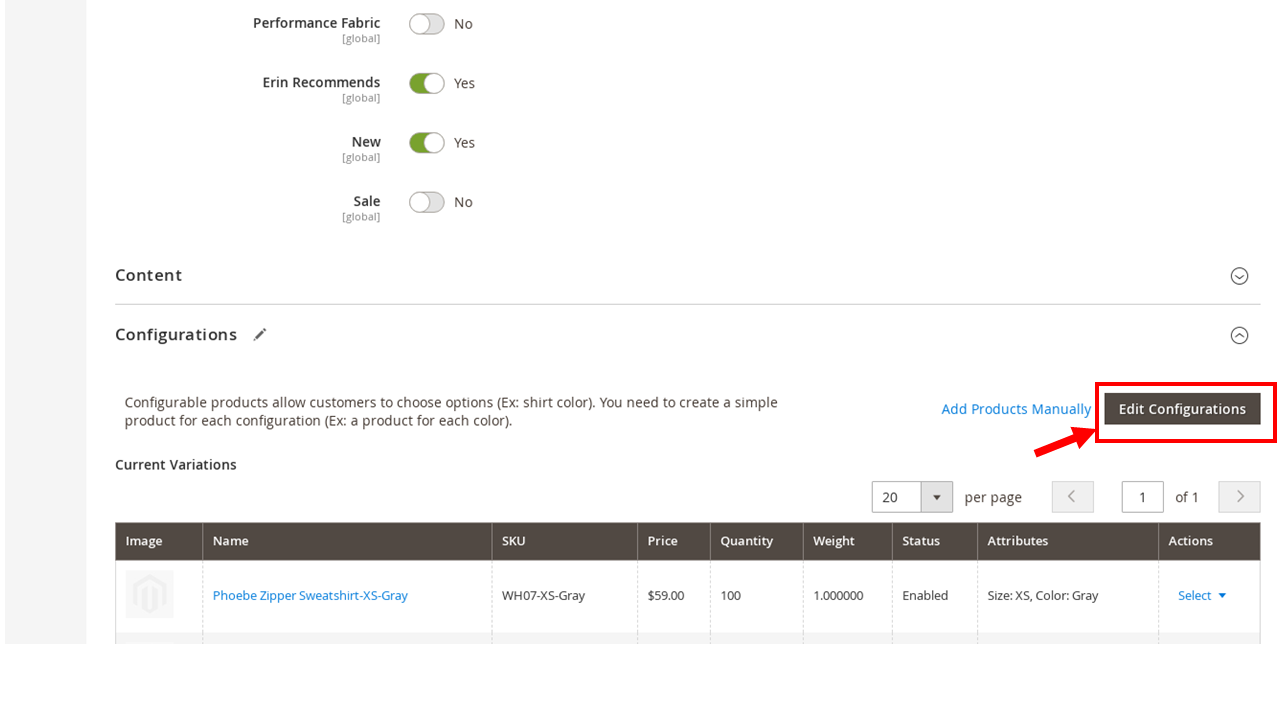} & <think>
The goal is to add a new color option Brown to size S for the configurable product ``Phoebe Zipper Sweatshirt.'' I am already on the product's edit page, and I can see the Configurations section with the ``Edit Configurations'' button. Since Size and Color are the configurable attributes (their comboboxes are disabled, indicating they're tied to configurations), \textbf{the correct workflow is to open the configurations wizard via ``Edit Configurations.''} In that wizard, I will select Size=S and Color=Brown, then proceed through the steps to generate the new simple product variant. The Brown option already exists in the Color attribute list (observed in the disabled combobox options), so I do not need to add a new attribute value in STORES. \textbf{The next step is to click ``Edit Configurations'' to start the wizard.}
</think> & Starting Wizard. \\
\midrule
\includegraphics[width=\linewidth]{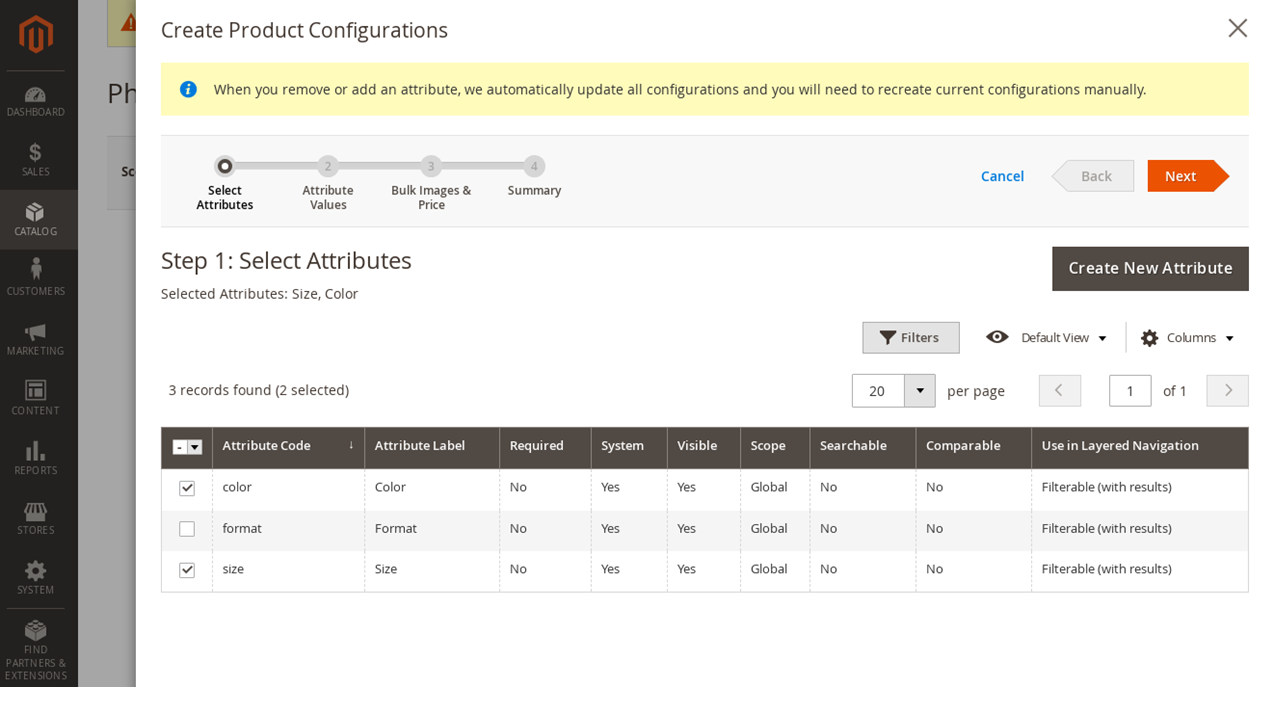} & <think>
The user wants to add a new Color option Brown specifically to Size S for the configurable product. I have the ``Create Product Configurations'' modal open on Step 1: Select Attributes, with Size and Color already selected, which matches the attributes we need. To proceed to choosing specific attribute values (where I can select Size S and Color Brown only), I need to go to Step 2: Attribute Values. \textbf{The ``Next'' button is present. Clicking it should take me to the values selection screen.}
</think> & Wizard Step 1. \\
\midrule
\multicolumn{3}{c}{\dots} \\
\midrule
\includegraphics[width=\linewidth]{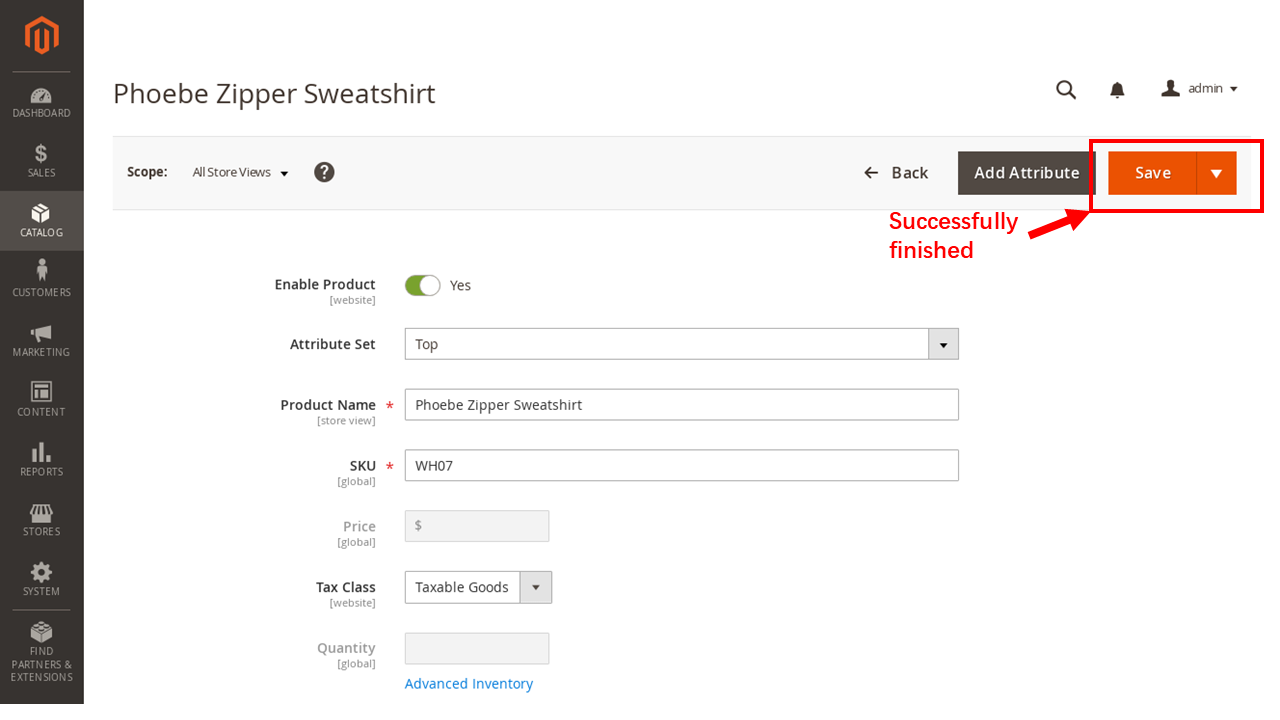} & <think>
The user's goal is to add a new color option "Brown" to Size "S" for the product "Phoebe Zipper Sweatshirt." I’m on the product edit page for the configurable product, and the Configurations section shows the "Current Variations" grid with a row "Phoebe Zipper Sweatshirt-S-Brown," SKU "WH07-S-Brown," Status "Enabled," and Attributes "Size: S, Color: Brown." This indicates the S–Brown variant exists and is enabled. I already executed a Save action earlier, and the variant appears in the grid, which implies the configuration was successfully generated and saved. \textbf{The best next step is to confirm completion to the user, including the variant name and SKU, aligning with the requirement to provide a complete answer that stands alone}.
</think> & Task Completed. \\
\end{xltabular}

\twocolumn

\end{document}